\documentclass[fleqn,twoside]{article}%
\usepackage{amsmath}
\usepackage{amsfonts}
\usepackage{amssymb}
\usepackage{graphicx}%
\def\be{\begin{equation}}
\def\ee{\end{equation}}
\def\bi{\bibitem}
\begin{document}
\title{Quantum cosmology of a classically constrained nonsingular Universe.}
\author{Abhik Kumar Sanyal}
\maketitle
\begin{center}
Dept. of Physics, Jangipur College, Murshidabad, \noindent
West Bengal, India - 742213. \\
\end{center}
\noindent
\begin{abstract}
The quantum cosmological version of a nonsingular Universe presented by Mukhanov and Brandenberger in the early nineties has been developed and the Hamilton Jacobi equation has been found under semiclassical (WKB) approximation. It has been pointed out that, parameterization of classical trajectories with semiclassical time parameter, for such a classically constrained system, is a nontrivial task and requires Lagrangian formulation
rather than the Hamiltonian formalism.
\end{abstract}
\noindent
\section{Introduction}
It transpires from Hawking-Penrose energy condition that the singularities of the general theory of relativity are unavoidable classically. In the absence of a complete and satisfactory theory of quantum gravity, it is not clear what would be the nature or the fate of singularities in the quantum domain. However, it might be possible in principle to get rid of the singularities classically by imposing some quantum mechanical bounds on certain
quantities. At the singularity, some of the curvatures, eg., energy momentum and even the Riemann tensors diverge. Further, quantum field theory predicts that, the scattering cross sections become infinity when all radiative corrections are taken into account. Renormalization makes individual terms manageable, but the entire series diverges. So in order to get rid of the singularities, curvature invariant terms must be constrained to take some upper limits. If Planck's length $l_{pl}$ is assumed to be the fundamental length, below which no length is measurable, then from dimensional argument, curvature invariants must be bounded as, $|R| < l_{pl}^{-2},\;|R_{\mu\nu}R^{\mu\nu}| < l_{pl}^{-4}$, etc. However, curvature invariants are infinite in number and there is no guarantee that imposing constraint on some lower order curvature invariant terms, all the higher order terms will be bounded. Hence, `Limiting Curvature Hypothesis' (LCH) \cite{fmm} had been invoked. LCH states that a finite number of curvature invariants should be bounded and one of those, say $I_2$, should have the property that $I_2 = 0$ singles out a particular nonsingular solution as the only one. In the process all the infinite set of curvature invariants are automatically bounded. If the limiting space is de-Sitter, then for a isotropic Universe,
LCH implies that the initial and the final stages of cosmological evolution for a closed Universe will give de-Sitter phase, while baby Universes will arise at the centre of the Black-Hole and eventually gravitational collapse is avoided. \\
Since Weyl tensor vanishes for an isotropic Universe, so in its simplest form, LCH can be realized by constraining only a couple of lower order curvature invariants, viz., $I_1 =~ ^{4}R$ and $I_2 = \sqrt{4 R_{\mu\nu}R^{\mu\nu} - R^2}$. Thus in its simplest form, ie., for homogeneous and isotropic space-time, Mukhanov and Brandenberger \cite{mb} followed by Brandenberger, Mukhanov and Sornborger \cite{bms} had constructed the following action (1), which preserves LCH.

\[ S = -\frac{1}{16\pi G}\int[(1+\phi_{1})R -(\phi_{2}+ \sqrt 3 \phi_{1})\sqrt{(4R_{\mu\nu}R^{\mu\nu} - R^2}) + V_{1}(\phi_{1}) + V_{2}(\phi_{2})]\sqrt{- g}~d^4 x.\]

\noindent
Here, the scalar field $\phi_{2}$ has been introduced to obtain nonsingular solutions, while $\phi_{1}$ bounds the curvature invariants for these nonsingular solutions. The potentials $V_{1}(\phi_{1})$ and $V_{2}(\phi_{2})$ should be chosen in a manner such that $(i)$ for $\phi_{i} \ll 1$, the
leading terms in the action gives back Einstein's theory and $(ii)$ their asymptotic behaviour should be such that LCH may be realized, which requires $V_{1}(\phi_{1}) \rightarrow$ constant as $\phi_1 \rightarrow \infty$, and $V_{2}(\phi_{2}) \rightarrow 0$ as $\phi_2\rightarrow \infty$. For a flat Robertson-Walker metric, they (\cite{mb} $\&$ \cite{bms}) had obtained a first order differential equation containing fields and the potentials and had drawn a phase diagram in $\phi_{1} - \phi_{2}$ plane, for some particular choice of the potentials. The phase diagram indicates nonsingular solutions in all regions and de-Sitter phase in the asymptotic region. In a nut-shell, the results obtained \cite{rb} are the following. Firstly, all the
homogeneous and isotropic cosmological solutions are nonsingular. Next, Two dimensional cosmological and Black-Hole solutions are nonsingular. Finally, there are evidence that nonsingular solutions may exist for four-dimensional Black-Holes and homogeneous-anisotropic cosmological models. Further, LCH has also been applied to dilaton-cosmology \cite{rre} and as a result, a class of spacially flat bouncing cosmological solutions have emerged.\\
This construction is definitely appealing in itself, however, it also incorporates some additional features in quantum domain. In the path integral formulation of quantum gravity, one has to analytically continue the Lorentzian space of indefinite metric to the Euclidean space of positive definite metric in order to ensure convergence of the path integral. Nevertheless, since Euclidean action of the gravitational field is not bounded from below, the path integral never really converges. The action presented by Mukhanov and Brandenberger \cite{mb} and Brandenberger, Mukhanov and Sornborger \cite{bms} is bounded from below, and therefore, the corresponding path integral converges. In addition, since the initial and the final stages of cosmological evolution are de-Sitter, so one gets Lorentzian wormholes instead of Euclidean. Finally, as the authors had claimed that the theory can
incorporate matter field and also can be extended to include anisotropic models, so it appears that the model is of greater interest with a wide spectrum of its applicability.\\
In the theory under consideration, two scalar fields, one of which is non-dynamical, have been nonminimally coupled to gravity, containing higher order curvature invariant term. Thus, along with the lapse function $N(t)$, an additional Lagrange multiplier is required in connection with the non-dynamical Scalar field, leading to classically constrained gravity theory. Study of Quantum cosmological aspects of such a constrained theory
may reveal more important cosmological implications. eg., in a recent work, such an attempt was made by Gabadadze and Shang \cite{gs} for a different
classically constrained theory of gravity, which has been found to admit solutions, absent from general theory of relativity. However, the new
solutions have some wonderful features like,-the spatially flat de-Sitter Universe can be created from nothing, has boundaries, with vanishing total energy, etc. \cite{ggs}. So, it appears worth to study the quantum cosmological aspects of the theory under consideration. Therefore, the constrained Hamiltonian dynamics of the theory under consideration has been studied along the line of Dirac's algorithm to find a constraint free primary Hamiltonian. Despite the fact that the theory under consideration contains constraints in addition to the Hamiltonian constraint of the standard theory of gravity, no attempt has been made to find the true degrees of freedom and to find the expression for the Hamiltonian in the reduced phase space. Rather, it has been quantized in the same tune of Wheeler-deWitt equation, $\hat H|\Psi> = 0$. However, since the only justification for a quantized method lies in its success, so the corresponding semiclassical approximation has been treated here explicitly.\\
The main motivation of this work is to study the connection of the quantum-cosmological equations for the theory under investigation with the
Hamilton-Jacobi equations of the classical theory and in the process to point out the difficulty in finding the semiclassical time required to parameterize classical trajectories. It has been observed that it is nontrivial task to extract semiclassical time parameter for such a constrained system. It has been shown that the semiclassical time parameter extracted in the usual manner does not lead to classical constraint equation. Further, the semiclassical time parameter may be found through speculation, but it is complex and requires Wick rotation. It has also been shown that in order to parameterize classical trajectories by an unique real time parameter, it is required to analyze Lagrangian constrained dynamics rather than Hamiltonian.\\
In the following section we have written down the classical field equations corresponding to the action under consideration for isotropic and homogeneous Robertson-Walker space time. The Hamiltonian and the corresponding Wheeler-deWitt equation have been found in section 3, after analyzing the constraint of the theory. Semiclassical approximation has been made by expressing the wave function as, $\Psi = e^{\frac{i}{\hbar}S}$,
and then by expanding $S$ in the powers of $M$ instead of $\hbar$. In the process, Hamilton-Jacobi equation has been found in section 4. In section 5, the usual procedure to extract the semiclassical time parameter has been attempted. However, it fails to fulfill the basic requirement of reproducing classical (Hamiltonian) constraint equation. Other naive techniques to find the time parameter have also been explored. Section 6 is devoted to follow an involved technique to find the unique semiclassical time parameter, which is the main motivation of the present paper. In section 7, a source term has been incorporated in the action and the same procedure has been followed. In the process, the functional Schr\"{o}dinger equation has been found and the uniqueness of the semiclassical time parameter has been proved without ambiguity. The outcome of the present work has been summarized in section (8). Finally, we have added up an appendix in section 9, to remind the readers about the semiclassical approximation in quantum cosmology.
\section{Formalism}
The action presented by Mukhanov and Brandenberger \cite{mb},
corresponding to a homogeneous and isotropic nonsingular Universe,
for which Weyl tensor vanishes, is given by,
\be S = -\frac{1}{16\pi G}\int[(1+\phi_{1})R -(\phi_{2}+\sqrt 3\phi_{1})\sqrt{(4R_{\mu\nu}R^{\mu\nu} - R^2}) + V_{1}(\phi_{1}) +
V_{2}(\phi_{2})]\sqrt{- g}~d^4 x,\ee

\noindent
where, $V_{1}(\phi_{1})$ and $V_{2}(\phi_{2})$ are the potentials corresponding to the pair of apparently non-dynamical scalar fields $\phi_{1}$ and $\phi_{2}$ respectively. LCH may be realized from the above action (1), only for some appropriate choice of the potentials. Action (1) leads to
Einstein's theory at small curvature, provided, $V_i(\phi_i) \sim \phi_i^2$, at $|{\phi_i}| \ll 1,$ where, $i = 1,2$. Further, to realize LCH, first requirement is to bound the curvature and the next is to obtain nonsingular solutions. Curvature is bounded provided, $V_1(\phi_1) \sim \phi_1$ at $|\phi_1| \gg 1$ and de Sitter solution is obtained in the asymptotic region, provided, $V_2(\phi_2) \sim$ constant, at $|\phi_2| \gg 1$.\\
Now, taking Robertson-Walker line element,

\be ds^2 = - N(t)^2 dt^2 + a^2(t) [\frac{dr^2}{1-kr^2} + r^2 (d\theta^2 + sin^2 \theta d\phi^2)],\ee

\noindent
where $N(t)$ is the lapse function, the Ricci scalar is given by,

\[R = -\frac{6}{N^2}(\frac{\ddot a}{a} + \frac{\dot{a^2}}{a^2} + N^2\frac{k}{a^2} -\frac{\dot N}{N}\frac{\dot a}{a}),\]

\noindent
and the next higher order curvature invariant term is,

\[\sqrt{4R_{\mu\nu}R^{\mu\nu} - R^2} = -\frac{\sqrt {12}}{N^2}(\frac{\ddot a}{a} - \frac{\dot a^2}{a^2} - N^2\frac{k}{a^2} - \frac{\dot N}{N}\frac{\dot a}{a}).\]

\noindent
So the action takes the following form,

\be S = \frac{\pi}{8G}\int[\frac{1}{N}(12\phi_{1} + 3\sqrt {12}\phi_{2} - 6)a\dot a^2 + \frac{\sqrt {12}}{N} a^2 \dot a \dot\phi_{2} + N(12\phi_{1}
+ \sqrt {12} \phi_{2} + 6)k a - N(V_{1} + V_{2})a^3]dt + S_{1},\ee

\noindent
where, the surface term is

\[S_{1} = - \frac{1}{8\pi G}\int_{\sigma}(1 - \frac{\phi_{2}}{\sqrt 3})\sqrt h~K~d^3 x,\]

\noindent
$K = \frac{1}{2N}\dot g_{ij} = -\frac{3}{N}\frac{\dot a}{a}$,
being the trace of the extrinsic curvature. The surface term is
clearly different from the one that appears in Einstein-Hilbert
action. However, the two are the same, if $\phi_{2}$ vanishes,
which has got a kinetic term in the action, and acts as a
dynamical variable. It is important to notice that the surface
term is not affected by the presence of the scalar field $\phi_1$,
since it acts as Lagrange multiplier only. In the above action $N$
and $\phi_{1}$ act as Lagrange multipliers, resulting in a pair of
constraint equations. Variation with respect to $N$ gives the
Hamiltonian constraint equation, but one can fix the gauge, $N =
1$, without loss of generality. However, we are not going to fix
the other Lagrange multiplier $\phi_{1}$, since it has been
invoked to bound the curvature invariants for the nonsingular
universe guaranteed by the field variable $\phi_{2}$. So,
altogether we get four equations (not all independent), two of
which are constraint equations and those obtained under the
variation with respect to $\phi_{2}$ and the scale factor $a$ are
the two field equations, where we set, $N = 1$, at the end. The
constraint equation appearing under the variation of $\phi_{1}$
is, \be \frac{\dot a^2}{a^2} + \frac{k}{a^2} = \frac{V_{1}'}{12},
\ee while the $\phi_{2}$ variation equation is, \be \frac{\ddot
a}{a} - \frac{\dot a^2}{a^2} - \frac{k}{a^2} =
-\frac{V_{2}'}{\sqrt {12}}. \ee The Hamiltonian constraint
equation, obtained by varying the action (3) with respect to $N$
is, \be (6-12\phi_{1}-3\sqrt{12}\phi_{2})\frac{\dot a^2}{a^2} -
\sqrt{12}\dot\phi_{2}\frac{\dot a}{a} + (6 + 12\phi_{1} +
\sqrt{12}\phi_{2})\frac{k}{a^2} - (V_{1} + V_{2})  = 0. \ee
Finally, varying the action with respect to the scale factor $a$,
one gets, \be 6(2\phi_{1} + \sqrt 3\phi_{2} - 1)\frac{\ddot a}{a}
+ 6(2\dot\phi_{1} + \sqrt 3\dot\phi_{2})\frac{\dot a}{a} +
3(2\phi_{1} + \sqrt 3\phi_{2} - 1)\frac{\dot a^2}{a^2} +\sqrt
3\ddot\phi_{2} - (6\phi_{1} + \sqrt 3\phi_{2}
+3)\frac{k}{a^2}+\frac{3}{2}(V_{1}+V_{2}) = 0. \ee In the above
field equations $V_{1}'(\phi_1)$ and $V_{2}'(\phi_2)$ denote the
derivatives of the potentials with respect to $\phi_{1}$ and
$\phi_{2}$ respectively. For spatially flat, $k = 0$ case, the
above field equations can be combined to yield a first order
differential equation, \be \frac{d\phi_{2}}{d\phi_{1}} =
\frac{V_{1}''}{V_{1}' V_{2}'}[-\frac{1}{4}(1-2\phi_{1})V_{1}' +
\frac{1}{2}(V_{1} + V_{2}) + \frac{3}{2\sqrt {12}}V_{1}'\phi_{2}].
\ee

\noindent
The phase diagram for a particular choice of the potentials was plotted by Mukhanov and Brandenberger \cite{mb}, showing four different classes of trajectories. In the first, the trajectory starts from the de-Sitter phase as $\phi_2 \rightarrow -\infty$ and evolves through to de-Sitter as $\phi_2 \rightarrow \infty.$ In the second, for small initial values of $\phi_1$ trajectory starts at $\phi_2 = -\infty$ reaches a turning point and returns
to $\phi_2 = -\infty$. In the third, the trajectory shows periodic solutions about Minkowski space-time $\phi_1 = \phi_2 = 0$. Finally, in the fourth class, trajectories starting with small $\phi_1$ and $\frac{\phi_1}{\phi_2}$, along with $\phi_2 \geq 0$, evolve towards de-Sitter at $\phi_2 = \infty$. In a nutshell, all the phase trajectories are either periodic about Minkowski space-time or else they asymptotically approach to de Sitter space. Hence all solutions are nonsingular. It is important to note that if $V_{1}''$ vanishes, $\phi_{2}$ turns out to be a constant and as a result $V_{2}'$ vanishes and the resulting solutions again become singular. However, the condition $V_{1}'' > 0$ is satisfied by all the potentials chosen by Mukhanov and Brandenberger
\cite{mb}.

\section{Analyzing the constraint and the Wheeler-deWitt equation}
It is important to note that in the process of developing an
action that might produce nonsingular cosmological solutions, the
determinant of the Hessian, $W_{ij} = \sum \frac{\partial^2
L}{\partial \dot q_{i}\partial \dot q_{j}}$, corresponding to the
Lagrangian of the above action (3) vanishes, and so the action
becomes singular. Vanishing of the determinant of Hessian, signals
the presence of constraint in the theory which should be analyzed
step by step carefully. Dirac algorithm \cite{ks} is the best
known technique to handle such constrained system and to construct
the Hamiltonian. Before we proceed, let us recapitulate a few important
artefact of Dirac's algorithm.\\
1. Firstly we remember that the existence of an infinite
invariance group leads to first class constraints, while singular
Lagrangians which do not possess a local gauge invariance leads to
second class constraints.\\
2. The rank of $W_{ij}$ for a system having $2N$ phase space
variables (that does not contain time explicitly) is $N$. For a
singular Lagrangian it is $R < N$. under this circumstances there
exists a non-degenerate $R \times R$ matrix $W_{\alpha a}$, such
that $p_i = \frac{\partial L}{\partial \dot q^i}$ can be solved
for $\dot q_a$ as $\dot q_a = f^a (q, p_a, \dot q^{\rho})$, where,
$a \longrightarrow 1, ...., R, \rho \longrightarrow R + 1, ....,N$.
In the process one finds, $r = N - R$ primary constraints, $p_r =
g_r (q, p_{\alpha})$, which originate from the definition of momentum.\\
3. Now if a function $F(q, p)$ be defined on the primary phase space
(this will be defined shortly) $\Gamma_p < \Gamma$, then the restriction
on $F$ to $\Gamma_p$ is achieved by replacing $p_r$ by $g_r (q, p_{\alpha})$.
If $F$ vanishes identically after this replacement, then it is called
weakly vanishing and is denoted by $F \approx 0$. If in addition the
gradient of $F$ also vanishes, then it is called strongly vanishing
and is denoted by $F \simeq 0$.\\
4. Primary constraints are expressed as $\phi_r(q, p) \approx 0,$
and the constrained Hamiltonian, defined on the constrained
phase space $\Gamma_c$, as,
\[H_c = \sum p_i\dot q_i -L.\]
One can write down the primary Hamiltonian defined on the primary phase space $\Gamma_p$, as,
\[H_p = H_c + \lambda^r\phi_r,\]
where, $\lambda^r$ are the Lagrange multipliers.\\
5. Next one has to check if the constraints are preserved in time,
i.e., \[\dot \phi_r = \{\phi_r, H_p\}\approx \{\phi_r, H_c\} +
\lambda^s\{\phi_r, \phi_s\}\equiv 0.\] If the Poisson bracket of a
constraint vanishes with all other at least weakly, then
it turns out to be first class constraint. In that case, $\{\phi_j, H_c\} \approx 0$,
and all these first class constraints are exhausted. The constraints $\phi_i$
which do not vanish modulo the constraint, leads to secondary constraints $\chi_k$.\\
6. One needs to check the consistency condition of the secondary
constraints again, i.e., if the constraints are preserved in time. Again for
the secondary constraints, if the poisson brackets of some (say $l$), vanish with all other,
then they are first class and for them $\{\chi_l, H_c\} \approx 0$.
Otherwise they are second class in nature and the Lagrange multipliers are
determined in the process. There are as many undetermined Lagrange multipliers
as there are first class primary constraints. If all the constraints are
second class then the primary Hamiltonian together
with the primary and secondary constraints constitute the field
equations, which may be solved in principle, leaving no
arbitrariness in the solutions.\\
7. Finally, one can try to find the true degrees of freedom and express the Hamiltonian in the reduced phase space $\Gamma_R$. However, it is a very difficult tusk, if not impossible in most of the situations. Even without finding the reduced Hamiltonian it is possible to quantize the theory. There is no unique prescription in this regard. Any quantum theory may be considered to be the correct one if it can go over to the classical counterpart through a suitable correspondence principle.\\
With this background knowledge, let us proceed to construct the
Hamiltonian of the theory. We understand that in order to construct
the primary Hamiltonian, one has to introduce all the constraints of
the theory in the Hamiltonian through Lagrange multiplier. Choosing
$\frac{\pi}{8G} = M$, $M$ being the square of the Planck mass, one
finds in view of the action (3) under the gauge choice $N = 1$,
\[\dot a = \frac{1}{M}\frac{p_{\phi_{2}}}{\sqrt {12} a^2},\;\;
\dot \phi_{2} = \frac{1}{M}\left(\frac{p_{a}}{\sqrt {12} a^2} -
\frac{2\phi_{1} + \sqrt 3\phi_{2} - 1}{a^3}p_{\phi_{2}}\right),\]
while $\dot\phi_{1}$ is not invertible due to the constraint, \be
C_{1} = p_{\phi_{1}} \approx 0.\ee
Since, $\frac{\partial C_1}{\partial p_{\phi_1}} \neq 0,$ so the
above constraint vanishes weakly. Hence the primary Hamiltonian
may be expressed as,
\[H_{p_{1}} = H_c + \alpha p_{\phi_{1}} = \sum p_i \dot q_i - L + \alpha p_{\phi_{1}}\]
\[=\frac{1}{M}\left(\frac{p_{a}p_{\phi_{2}}}{\sqrt {12} a^2}
- \frac{2\phi_{1} + \sqrt 3\phi_{2} -
1}{2a^3}p_{\phi_{2}}^2\right) - M[(12\phi_{1} + \sqrt {12}
\phi_{2} + 6)ka - (V_{1} + V_{2})a^3] + \alpha p_{\phi_{1}} = 0,\]
where, $\alpha$ is the Lagrange multiplier. Since, $\{C_1, H_c\}$ does
not vanish even weakly, so $C_{1}$ is a second class primary constraint.
As there are no first class primary constraint, so there will be no
undetermined Lagrange multiplier of the theory. Now the constraint must be preserved in time, i.e.,
\be \dot C_{1} = \{C_{1}, H_{p_{1}}\} = D_{1}=
\frac{1}{M}\frac{p_{\phi_{2}}^2}{a^3} + M(12ka - V_{1}' a^3)
 \approx 0. \ee  This is a new second class constraint
as long as $V_{1}''$ exists, since $\{C_{1}, \dot C_{1}\} =  \{C_1, D_1\} = M
V_{1}'' a^3$. As already pointed out that for the existence of
nonsingular solutions $V_{1}''$ must not vanish, here we again
observe that $V_{1}''$ must not vanish to remove arbitrariness
from the primary Hamiltonian. However no such restriction is
required for $V_{2}$. Again the condition that the constraint should be preserved in time leads to,
\be
\dot{D_{1}} = \{D_{1}, H_{p_{1}}\} = -2V_{2}' p_{\phi_{2}} - M
\alpha a^3 V_{1}'' \approx 0, \ee modulo the constraint (10).
This is not a new constraint,
rather it fixes the Lagrange multiplier $\alpha = -\frac{2 V_2 '
p_{\phi_2}}{M a^3 V_1 ''}$, provided $V_{1}''$ exists. Thus both
the primary and the secondary constraints are second class. Hence
the primary Hamiltonian, being free from arbitrariness can now be
expressed as, \be H_{p_{1}} =
\frac{1}{M}\left(\frac{p_{a}p_{\phi_{2}}}{\sqrt {12} a^2}  -
\frac{2\phi_{1} + \sqrt 3\phi_{2} - 1}{2a^3}p_{\phi_{2}}^2
-\frac{2V_{2}'}{a^3 V_{1}''}p_{\phi_{1}}p_{\phi_{2}}\right) -
M\left[(12\phi_{1} + \sqrt {12} \phi_{2} + 6)ka - (V_{1} +
V_{2})a^3\right] = 0. \ee

\noindent
One can now easily check that we have obtained the correct Hamiltonian in the usual manner. The only exception appears to be with
$\dot \phi_{1} = \frac{\partial H}{\partial p_{\phi_{1}}} = -\frac{2V_{2}'}{Ma^3 V_{1}''}p_{\phi_{2}}$, which looks like a new equation. However, this is the one, we were in search of, since $\dot\phi_{1}$ can now be inverted through this equation. In any case, it is not an independent equation, since taking time derivative of the first (constraint) equation (4) and using the second one (5), one can arrive at it. Thus the Hamiltonian, (12) being free from constraints and being able to produce all the field equations, is the correct one. To make a comparison, let us take Einstein-Hilbert action minimal coupled to a scalar field,

\[A = 12M\int[-\frac{1}{2}a\dot a^2 +k\frac{a}{2}+\frac{1}{12M}(\frac{1}{2}\dot\phi^2 -V(\phi))a^3]dt,\]
and write down the corresponding Hamiltonian,

\be
-\frac{1}{24M}\frac{p_{a}^2}{a}+\frac{p_{\phi}^2}{2 a^3} - 6M k a
+ a^3 V(\phi) = 0.\ee

\noindent
It is interesting to note that the classical field equations (4) through (7) reduce to the vacuum Einstein's equations at any stage of cosmic evolution as $\phi_{1} = \phi_{2} = 0$. However, the primary Hamiltonian (12) once constructed in view of the constrained system under investigation, is different from (13) and does not ever reduce to the vacuum Einstein's equation. This fact has been manifested in the linear appearance of $p_{a}$ in $H_{p_{1}}$. The fact that here $p_{a}$ appears linearly in the Hamiltonian, clearly differentiates all actions (corresponding to minimal and standard non-minimal coupling) with the present one.\\

\noindent
Our next attempt will be to canonically quantize the theory under consideration. There is no standard prescription to quantize a classically constrained system and the only justification of a particular method of quantization lies in its success. The meaning of the last sentence is that one should be able to find a correspondence principle to go over to the classical theory under semiclassical limit. We shall write down the counterpart of the Wheeler-deWitt equation $\hat H|\Psi> = 0$, for the modified theory of gravity under consideration and show in the following sections how to find the notion of semiclassical time so that classical field equations may be reproduced.
Now, in constructing the Wheeler-deWitt equation corresponding to the Hamiltonian (12), operator ordering ambiguities should be taken care off. Some of the operator ordering ambiguities may be removed by expressing, $\hat{p_{a}} \rightarrow q^{-1}(a)\hat{p_{a}}q(a)$, where, $q(a)$ is an arbitrary function of $a$. Hence the first term of the Wheeler-deWitt equation corresponding to the Hamiltonian (12), after replacing $\hat p$ by $-i \hbar\nabla$, turns out to be
$$-\frac{1}{\sqrt{12}a^2}[\frac{\partial^2 \Psi}{\partial a\partial\phi_{2}}
+ \frac{1}{q}\frac{\partial q }{\partial a}\frac{\partial
\Psi}{\partial\phi_{2}}].$$

\noindent
Clearly, first order derivative of the
wavefunction $\Psi$ with respect to $a$, (ie., $\frac{\partial
\Psi}{\partial a}$) does not appear, while the same with respect
to the scalar field (ie., $\frac{\partial \Psi}{\partial
\phi_{2}}$) appears. Further, operator ordering in $\hat
p_{\phi_{1}}$ and $\hat p_{\phi_{2}}$ appearing due to the
presence of the second and the third terms in (12), again introduces
first derivative terms (ie., $\frac{\partial \Psi}{\partial
\phi_{1}}$ and $\frac{\partial \Psi}{\partial \phi_{2}}$) in the
Wheeler-deWitt equation, corresponding to the scalars $\phi_i$.
However, from the regularity argument \cite{sdc}, it follows that
$\frac{\partial \Psi}{\partial \phi_{i}}$ can be neglected at
sufficiently small values of $a$, ie, at sufficiently early epoch.
Hence, the Wheeler-deWitt equation may be kept free from the
first derivative terms even after removing some of the operator
ordering ambiguities (unlike the situation encountered in standard
and other nonstandard theories of gravity), and is expressed as,
\be \left[\frac{\hbar
^2}{M}\left\{\frac{1}{\sqrt{12}a^2}\frac{\partial^2}{\partial
a\partial \phi_{2}}-\frac{2\phi_{1}+\sqrt 3 \phi_{2}-1}{2 a^3}
\frac{\partial^2}{\partial \phi_{2}^2}-\frac{2V_{2}'}{a^3
V_{1}''}\frac{\partial^2}{\partial
\phi_{1}\partial\phi_{2}}\right\}
+M\left\{(12\phi_{1}+\sqrt{12}\phi_{2}+6)ka -
(V_{1}+V_{2})a^3\right\}\right]|\Psi> = 0, \ee which is
independent of the operator ordering parameter $q(a)$. Not all quantum states $|\Psi>$ of the Wheeler -deWitt equation (14) are allowed, since they are constrained by the quantum analogue of the classical constraints (10) and (11). However, since our aim is not to find the solution of (14), so we neither, are in search of true degrees of freedom nor incorporate the quantum analogue of the classical constraints. Rather, for the sake of comparison,
we write down the Wheeler-deWitt equation corresponding to Einstein's
gravity with a minimally coupled scalar field (see Appendix),
\[\left[\frac{\hbar^2}{24M}(\frac{\partial^2}{\partial
a^2}+\frac{q}{a}\frac{\partial}{\partial
a}-\frac{12M}{a^2}\frac{\partial^2}{\partial\phi^2})-6Mk
a^2 +a^4 V(\phi)\right]|\Psi> = 0,\] where, $q$ takes care of
some of the operator ordering ambiguities. We observe that, neither
$\frac{\partial^2}{\partial a^2}$, nor $\frac{\partial}{\partial a}$ term appears in the Wheeler-deWitt equation (14), which may have some deep significance, not presently known.
\section{Semiclasical approximation}
Reparametrization invariance of the theory of gravity leads to the
Hamiltonian constraint yielding Wheeler-deWitt equation. Despite the fact that the theory under consideration has additional constraints, we have not made any attempt to find the true degrees of freedom to construct the Hamiltonian in the reduced phase space $\Gamma_R$. Rather, the corresponding quantum equation (14) has been constructed in the same tune of the standard Wheeler-deWitt equation $\hat{H}|\Psi> = 0$. Now the obvious question is, "does equation (14) alone represents the quantum version of the classical field equations (4) through (7)"? Other way round one may ask, "is it possible to set up a correspondence between the Wheeler-deWitt equation (14) and the classical equations (4) through (7)"?. Since, the constraint free Hamiltonian represents the correct one to reproduce all the classical field equations, so the answer to the question raised above is positive, if under a suitable semiclassical prescription, one can find a semiclassical notion of time to recover the Hamilton constraint equation (6) from the Wheeler-deWitt equation (14). It is a nontrivial task which we shall take up in this and in the following sections.\\
At energy below Planck scale, the wave function can be expressed as,
$\Psi(a, \phi_1, \phi_2) = e^{\frac{i}{\hbar}S(a, \phi_1, \phi_2)}$.
Expanding, $S$ in the powers of $M$ as,
\[S = M S_{0} + S_{1} + M^{-1} S_{2} + \cdots ,\]
and inserting it in the Wheeler-deWitt equation (14), one
obtains
\[\frac{\hbar^2}{M}[-\frac{1}{\sqrt{12}a^2}
\left\{\frac{i}{\hbar}\frac{\partial^2(M S_{0} + S_{1}
+ M^{-1} S_{2} + \cdots)}{\partial a\partial\phi_{2}}
- \frac{1}{\hbar^2}\frac{\partial(M S_{0} + S_{1}
+ M^{-1} S_{2} + \cdots)}{\partial a}\frac{\partial(M S_{0}
+ S_{1} + M^{-1} S_{2} + \cdots)}{\partial\phi_{2}}\right\}\]
\[+ \frac{2\phi_{1}+\sqrt3 \phi_{2}-1}{2a^3}
\left\{\frac{i}{\hbar}\frac{\partial^2(M S_{0} + S_{1}
+ M^{-1} S_{2} + \cdots)}{\partial \phi_{2}^2}
- \frac{1}{\hbar^2}\left(\frac{\partial(M S_{0} + S_{1}
+ M^{-1} S_{2} + \cdots)}{\partial\phi_{2}}\right)^2\right\}\]
\[+ \frac{2V_{2}'}{a^3 V_{1}''}\left\{\frac{i}{\hbar}
\frac{\partial^2(M S_{0} + S_{1} + M^{-1} S_{2} + \cdots)}
{\partial a\partial \phi_{2}} - \frac{1}{\hbar^2} \frac{\partial(M
S_{0} + S_{1} + M^{-1} S_{2} + \cdots)}
{\partial\phi_{1}}\frac{\partial(M S_{0} + S_{1} + M^{-1} S_{2} +
\cdots)}{\partial\phi_{2}}\right\}]\] \be - M[(12\phi_{1} +
\sqrt{12}\phi_{2} + 6) -(V_{1} + V_{2})a^3] = 0.\ee Let us now
collect expression having same powers in $M$. For Einstein-Hilbert
action with minimally coupled fields one gets to the power of
$M^2$ an expression that states that the Hamilton-Jacobi function
$S_{0}$ depends on three space only (see Appendix). Here the situation
is quite different in the sense that we do not get expressions
corresponding to the order $M^2$. To the order $M^1$, we have,
\be\frac{1}{\sqrt{12}a^2}\frac{\partial S_{0}}{\partial
a}\frac{\partial S_{0}}{\partial \phi_{2}} - \frac{2\phi_{1} +
\sqrt 3\phi_{2} -1}{2a^3}\left(\frac{\partial
S_{0}}{\partial\phi_{2}}\right)^2 - \frac{2V_{2}'}{a^3
V_{1}''}\frac{\partial S_{0}}{\partial\phi_{1}}\frac{\partial
S_{0}}{\partial\phi_{2}} - (12\phi_{1}+\sqrt{12}\phi_{2}+6)ka +
(V_{1}+V_{2})a^3 = 0.\ee This is the Hamilton-Jacobi equation. It can
be identified with the Hamiltonian  constraint equation (6) only under
an appropriate choice of semiclassical time parameter. Following two
sections are devoted to find the semiclassical time parameter.
\section{Semiclassical time parameter - standard technique}
Since in view of the Hamiltonian (12) all the velocities are now invertible and
so following standard technique as in Einstein's gravity with minimally
coupled scalar field (see Appendix, 9.2), the time parameter can be found as,
\[\frac{\partial}{\partial t} = \frac{1}{\sqrt{12}a^2}\frac{\partial S_{0}}
{\partial\phi_{2}}\frac{\partial}{\partial a} - \frac{2M
V_{2}'}{V_{1}'' a^3} \frac{\partial
S_{0}}{\partial\phi_{2}}\frac{\partial}{\partial\phi_{1}} +
\left(\frac{1}{\sqrt{12}a^2}\frac{\partial S_{0}}{\partial a} -
(\frac{2\phi_{1}+\sqrt 3\phi_{2}-1}{a^3})\frac{\partial
S_{0}}{\partial \phi_{2}}\right)
\frac{\partial}{\partial\phi_{2}}.\] The problem is that, this
choice of time parameter does not lead to classical constraint
equation (6). This is because, $\frac{\partial
S_{0}}{\partial\phi_{1}}$ appearing in the Hamilton-Jacobi
equation (16) is not obtainable from it and as a result, remains
arbitrary. This is the source of trouble that we encounter in the present situation,
to parameterize classical trajectories with semiclassical time
parameter following usual procedure. Thus, the standard procedure
does not work in the classically constrained system under consideration.\\
One can try to find the same by expressing
the time parameter as
\[\frac{\partial}{\partial t} = \left(b\frac{\partial S_{0}}{\partial a}
+  c\frac{\partial S_{0}}{\partial\phi_{1}}+d\frac{\partial
S_{0}}{\partial\phi_{2}}\right) \frac{\partial}{\partial
a}+\left(j\frac{\partial S_{0}}{\partial a} + l\frac{\partial
S_{0}}{\partial\phi_{1}}+m\frac{\partial
S_{0}}{\partial\phi_{2}}\right)\frac{\partial}{\partial\phi_{1}}+\left(u\frac{\partial
S_{0}}{\partial a}+v\frac{\partial
S_{0}}{\partial\phi_{1}}+n\frac{\partial
S_{0}}{\partial\phi_{2}}\right)\frac{\partial
}{\partial\phi_{2}},\] where, $b$, $c$, $d$, $j$, $l$, $m$, $u$,
$v$ and $n$ are arbitrary functions of $a, \phi_{1}$ and
$\phi_{2}$. It is now possible to find $\dot a, \dot\phi_{1}$ and
$\dot\phi_{2}$, using this expression as,
\[\dot a = b\frac{\partial S_{0}}{\partial a}+c\frac{\partial S_{0}}
{\partial\phi_{1}}+d\frac{\partial S_{0}}{\partial\phi_{2}},\]
\[\dot\phi_{1} = j\frac{\partial S_{0}}{\partial a}+l\frac{\partial S_{0}}
{\partial\phi_{1}}+m\frac{\partial S_{0}}{\partial\phi_{2}},\]
\[\dot\phi_{2} = u\frac{\partial S_{0}}{\partial a}+v\frac{\partial S_{0}}
{\partial\phi_{1}}+n\frac{\partial S_{0}}{\partial\phi_{2}}.\]
Substituting all these expressions in equation (6) and equating
the coefficients of $\nabla S_{0}$ etc., between the equation thus
formed and the Hamilton-Jacobi equation (16), one can finally
arrive at the following time parameter,
\[\frac{1}{n}\frac{\partial}{\partial t} = \left\{-\frac{\sqrt{12}a}
{6(2\phi_{1}+\sqrt 3 \Phi_{2}-1)}\frac{\partial S_{0}}{\partial a}
+\frac{4V_{2}'}{(2\phi_{1}+\sqrt 3
\phi_{2}-1)V_{1}''}\frac{\partial S_{0}}
{\partial\phi_{1}}+\frac{\partial S_{0}}{\partial
\phi_{2}}\right\}\frac{\partial} {\partial
a}\]\[+\left\{\frac{\partial S_{0}}{\partial a}
-\frac{2\sqrt{12}V_{2}'}{aV_{1}''}\frac{\partial
S_{0}}{\partial\phi_{1}} -\frac{\sqrt 3}{(2\phi_{1}+\sqrt 3
\phi_{2}-1)a} \left(1+\frac{1}{12n^2 a^4}\right)\frac{\partial
S_{0}}{\partial\phi_{2}}\right\} \frac{\partial}{\partial
\phi_{2}}.\] With this technique of parametrization, classical
constraint equation (6) is automatically reproduced, but the problem
associated with this time parameter is that, there still exists an
arbitrariness in the form of the arbitrary parameter $n = n(a, \phi_1, \phi_2)$. Thus,
the time parameter is not unique, and something else should be tried.\\
It is interesting to note
that simply by inspection one can choose a time parameter free
from such arbitrariness as,
\[\frac{\partial}{\partial t}=i\left[\frac{1}{\sqrt{12}a^2}\frac{\partial S_{0}}
{\partial\phi_{2}}\frac{\partial}{\partial a}
-\frac{1}{\sqrt{12}a^2} \frac{\partial S_{0}}{\partial
a}\frac{\partial}{\partial\phi_{2}} +\frac{2V_{2}'}{a^3
V_{1}''}\frac{\partial S_{0}}{\partial a}
\frac{\partial}{\partial\phi_{2}}\right].\] This time parameter
reproduces equation (6), in view of Hamilton-Jacobi equation (16).
However, this is purely intuitive on one hand and is imaginary on
the other. So we must find an involved technique for this purpose.
\section{Semiclassical time parameter - an involved technique}
As already mentioned, equation (4) is an additional gravitational
constraint equation of the theory under consideration, since it
does not contain second order derivatives. The
standard technique should be to differentiate equation (4) and
then to compare it with equation (6). In the process, if the
emerging equation is again a constraint equation, then it should
be entered into the Lagrangian via Lagrange multiplier. The
emerging constraint equation in this case is, \be\dot\phi_{1} +
2\sqrt{12}\frac{V_{2}'}{V_{1}''}\frac{\dot a}{a} = 0.\ee Hence,
introducing this  constraint equation (17) in action (3) via a
Lagrange multiplier $\lambda$, we obtain, \be S =
M\int\left[(12\phi_{1}+3\sqrt{12}\phi_{2}-6)a\dot a^2 +
\sqrt{12}a^2\dot
a\dot\phi_{2}+(12\phi_{1}+\sqrt{12}\phi_{2}+6)ka-(V_{1}+V_{2})a^3
-\lambda(\dot\phi_{1}+2\sqrt{12}\frac{V_{2}'}{V_{1}''}\frac{\dot
a}{a})\right]dt.\ee The canonical momenta are found from the
action (18) as, \be
p_{a}=M[2(12\phi_{1}+3\sqrt{12}\phi_{2}-6)a\dot
a+\sqrt{12}a^2\dot\phi_{2}-2\sqrt{12}\lambda\frac{V_{2}'}{aV_{1}''}]\;;\;\;p_{\phi_{1}}
= -M\lambda\;;\;\;p_{\phi_{2}} =\sqrt{12}Ma^2\dot a\;.\ee We don't
write down the field equations since we are not going for
classical solutions. However, it is important to note that since
variation with respect to $\lambda$ gives back the constraint
equation (17) and in view of (19) the classical constraint
equation (6) remains unchanged, so, the primary Hamiltonian (12),
the Wheeler-deWitt equation (14) and the Hamilton-Jacobi equation
(16) are systematically reproduced. Further, identifying canonical
momenta with corresponding derivatives of the Hamilton-Jacobi
function, equation (12) can be found again from Hamilton-Jacobi
equation (16). Now in order to obtain equation (6) from (16), let
us use equations (17) and (19), which gives, \be \dot a =
\frac{p_{\phi_{2}}}{M\sqrt{12}a^2} =
\frac{1}{\sqrt{12}a^2}(\frac{\partial
S_{0}}{\partial\phi_{2}}),\;\;ie.,\;\; \frac{\partial}{\partial
t}|_{a} = \frac{1}{\sqrt{12}a^2}(\frac{\partial
S_{0}}{\partial\phi_{2}})\frac{\partial}{\partial a},\ee \be
\dot\phi_{1} = -\frac{2V_{2}'}{a^3
V_{1}''}\frac{p_{\phi_{2}}}{M},\;\;ie.,\;\;\frac{\partial}{\partial
t}|_{\phi_{1}} = -\frac{2V_{2}'}{a^3 V_{1}''}(\frac{\partial
S_{0}}{\partial\phi_{2}})\frac{\partial}{\partial\phi_{1}},\ee
\[ \dot\phi_{2} = \frac{1}{\sqrt{12a^2}}\frac{p_{a}}{M}
-\frac{2\phi_{1}+\sqrt 3\phi_{2}-1}{a^3}\frac{p_{\phi_{2}}}{M}
-\frac{2\sqrt{12}V_{2}'}{aV_{1}''}\frac{p_{\phi_{1}}}{M},\]\be
ie.,\;\;\frac{\partial}{\partial t}|_{\phi_{2}} =
\left(\frac{1}{\sqrt{12}a^2}(\frac{\partial S_{0}}{\partial a})
-\frac{2\phi_{1}+\sqrt 3\phi_{2}-1}{a^3}(\frac{\partial S_{0}}
{\partial\phi_{2}})-\frac{2\sqrt{12} V_{2}'}{aV_{1}''}
(\frac{\partial
S_{0}}{\partial\phi_{1}})\right)\frac{\partial}{\partial
\phi_{2}}.\ee Equations (20), (21) and (22) are now combined to
yield the correct and unique semiclassical time parameter, \be
\frac{\partial}{\partial t} =  \frac{1}{\sqrt{12}a^2}
(\frac{\partial S_{0}}{\partial\phi_{2}})\frac{\partial}{\partial
a} -\frac{2V_{2}'}{a^3 V_{1}''}(\frac{\partial
S_{0}}{\partial\phi_{2}})
\frac{\partial}{\partial\phi_{1}}+\left(\frac{1}{\sqrt{12}a^2}
(\frac{\partial S_{0}}{\partial a})-\frac{2\phi_{1}+\sqrt
3\phi_{2}-1}{a^3} (\frac{\partial
S_{0}}{\partial\phi_{2}})-\frac{2\sqrt{12} V_{2}'}{aV_{1}''}
(\frac{\partial
S_{0}}{\partial\phi_{1}})\right)\frac{\partial}{\partial
\phi_{2}}.\ee One can now easily obtain $\frac{\partial
S_{0}}{\partial a}, \frac{\partial S_{0}}{\partial \phi_{1}}$ and
$\frac{\partial S_{0}} {\partial \phi_{2}}$ from above time
parameter (23) and substituting these in equation  (16), one can
reproduce equation (6). To check whether we have found the correct
and unique semiclassical time parameter let us continue by including
a source term (in the form of a dynamical scalar field) in the model.
\section{Including a source term}
In this section we consider an additional source term in the
action (1), in the form of a minimally coupled scalar field with
Lagrangian density,
\[L_{m} = - \frac{1}{2\pi^2} [\frac{1}{2}
\sigma_{,\alpha}\sigma_{,\beta}g^{\alpha\beta} + U(\sigma)],\]
where, $\sigma$ is the scalar field and $U(\sigma)$ is an
arbitrary potential. The action (1) now reads (taking $M =
\frac{\pi}{8G}$, as before),
\be S = M\int\left[(12\phi_{1}
+ 3\sqrt {12}\phi_{2} - 6)\frac{a\dot a^2}{N} + \frac{\sqrt{12}}{N} a^2 \dot
a \dot\phi_{2} + N[(12\phi_{1} + \sqrt {12} \phi_{2} + 6) k a -
(V_{1} + V_{2})a^3] + \frac{a^3}{M}\left(\frac{\dot\sigma^2}{N}-
NU(\sigma)\right)\right]dt.\ee Corresponding field equations are (under
variation of the above action with respect to $\phi_{1}, \phi_{2},
\sigma, N$ and $a$ and setting $N = 1$),
\be \frac{\dot a^2}{a^2}
+ \frac{k}{a^2} = \frac{V_{1}'}{12}, \ee
\be \frac{\ddot a}{a} -
\frac{\dot a^2}{a^2} - \frac{k}{a^2} = -\frac{V_{2}'}{\sqrt {12}},
\ee
\be \ddot\sigma + 3\frac{\dot a}{a}\dot\sigma + U'(\sigma) =
0, \ee
\be M[(12\phi_{1}+3\sqrt{12}\phi_{2}-6)\frac{\dot a^2}{a^2}
+\sqrt{12}\dot\phi_{2}\frac{\dot a}{a} - (12\phi_{1} +
\sqrt{12}\phi_{2}+6)\frac{k}{a^2} + (V_{1} +
V_{2})]+\frac{\dot\sigma^2}{2}+U(\sigma)  = 0, \ee
\be 6(2\phi_{1}
+ \sqrt 3\phi_{2} - 1)\frac{\ddot a}{a} + 6(2\dot\phi_{1} + \sqrt
3\dot\phi_{2})\frac{\dot a}{a} + 3(2\phi_{1} + \sqrt 3\phi_{2} -
1)\frac{\dot a^2}{a^2} +\sqrt 3\ddot\phi_{2} - (6\phi_{1} + \sqrt
3\phi_{2} +3)\frac{k}{a^2}+\frac{3}{2}(V_{1}+V_{2})
+\frac{1}{2M}(\frac{\dot\sigma^2}{2} - U)= 0, \ee where,
$V_{1}'(\phi_{1}), V_{2}'(\phi_{2})$ and $U'(\sigma)$ are the
derivatives of the potentials with respect to $\phi_{1}, \phi_{2}$
and $\sigma$ respectively. As before we differentiate equation
(25) with respect to time and compare it with equation (26) to get
the constraint equation
\be \dot\phi_{1} +
2\sqrt{12}\frac{V_{2}'}{V_{1}''}\frac{\dot a}{a} = 0.\ee Next we
incorporate this equation (30) into the Lagrangian through the same
Lagrange multiplier $\lambda$, as before. In view of the Lagrangian
so formed, we obtain the same set of canonical momenta (19) along
with an additional one, corresponding to the field $\sigma$, viz.,
\be p_{\sigma} = a^3 \dot \sigma. \ee
Variation of the Lagrangian with respect to the Lagrange
multiplier $\lambda$ returns equation (30). So, in view of
equations (19), (30) and (31), one can express all the velocities
in terms of momenta, and thus the primary Hamiltonian free from
arbitrariness of Lagrange multiplier is found as,
\be H_{p} =
\frac{p_{a}p_{\phi_{2}}}{\sqrt{12}Ma^2}-\frac{2V_{2}'}{Ma^3V_{1}''}p_{\phi_{1}}p_{\phi_{2}}-\frac{2\phi_{1}+\sqrt
3\phi_{2}-1}{2Ma^3}p_{\phi_{2}^2}+\frac{p_{\sigma}^2}{2a^3}-M[(12\phi_{1}+\sqrt{12}\phi_{2}+6)ka
- (V_{1}+V_{2})a^3]+a^3U(\sigma) = 0. \ee Corresponding
Wheeler-deWitt equation $\hat{H_{p}}|\Psi > = 0$ is,
\[[\frac{\hbar
^2}{M}\left\{-\frac{1}{\sqrt{12}a^2}\frac{\partial^2}{\partial
a\partial \phi_{2}}+\frac{2\phi_{1}+\sqrt 3 \phi_{2}-1}{2 a^3}
\frac{\partial^2}{\partial \phi_{2}^2}+\frac{2V_{2}'}{a^3
V_{1}''}\frac{\partial^2}{\partial
\phi_{1}\partial\phi_{2}}\right\}\]\be-M\left\{(12\phi_{1}+\sqrt{12}\phi_{2}+6)ka -
(V_{1}+V_{2})a^3\right\}-\frac{\hbar^2}{2a^3}\frac{\partial^2}{\partial\sigma^2}+a^3
U(\sigma)\;]|\Psi> = 0.\ee As before, we proceed to make
semiclassical approximation by expressing
$\Psi[a,\phi_{1},\phi_{2},\sigma]$ as $\Psi =
\exp[\frac{i}{\hbar}S(a,\phi_{1},\phi_{2},\sigma)]$, expanding $S$
in power series of $M$ as, $S = MS_{0}+S_{1}+M^{-1}S_{2}+\cdots
etc.$, and then substituting all these in equation (33). Finally,
equating terms having different orders of $M$ to zero, we find for
$M^2$ order \be \frac{\partial S_{0}}{\partial \sigma} =
0,\;\;ie.,\;\;S_{0} = S_{0}(a,\phi_{1},\phi_{2}). \ee $M^1$ order
term leads to Hamilton-Jacobi equation for the source free
nonsingular gravitational field as,
\be
\frac{1}{\sqrt{12}a^2}\frac{\partial S_{0}}{\partial
a}\frac{\partial S_{0}}{\partial \phi_{2}}-\frac{(2\phi_{1}+\sqrt 3
\phi_{2}-1)}{2a^3}(\frac{\partial
S_{0}}{\partial\phi_{2}})^2-\frac{2V_{2}'}{a^3
V_{1}''}\frac{\partial S_{0}}{\partial\phi_{1}}\frac{\partial
S_{0}}{\partial\phi_{2}}-(12\phi_{1}+\sqrt{12}\phi_{2}+6)ka+(V_{1}+V_{2})a^3=0.
\ee
One can easily identify this equation with the Hamilton-Jacobi
equation (16). It leads to the classical source free gravitational
constraint equation (6) under the same choice of the time
parameter (23). Hence, source free Hamiltonian (12) can be found from it by
identifying the canonical momenta with corresponding derivatives
of $S_{0}$, as before. The next, ie., $M^0$ order term gives the
following equation,
\pagebreak
\[-\frac{1}{\sqrt{12}a^2}\left(i\hbar\frac{\partial^2 S_{0}}{\partial a\partial\phi_{2}}
-\frac{\partial S_{0}}{\partial a}\frac{\partial S_{1}}{\partial
\phi_{2}} -\frac{\partial S_{1}}{\partial a}\frac{\partial
S_{0}}{\partial \phi_{2}}\right) +\frac{2V_{2}'}{a^3
V_{1}''}\left(i\hbar\frac{\partial^2 S_{0}}
{\partial\phi_{1}\partial \phi_{2}}-\frac{\partial S_{0}}{\partial
\phi_{1}} \frac{\partial S_{1}}{\partial \phi_{2}}-\frac{\partial
S_{1}}{\partial \phi_{1}} \frac{\partial S_{0}}{\partial
\phi_{2}}\right)\]\be + \frac{2\phi_{1}+\sqrt
3\phi_{2}-1}{2a^3}\left(i\hbar\frac{\partial^2 S_{0}} {\partial
\phi_{2}^2}-2\frac{\partial S_{0}}{\partial
\phi_{2}}\frac{\partial S_{1}} {\partial
\phi_{2}}\right)-\frac{1}{2a^3}\left(i\hbar\frac{\partial^2 S_{1}}
{\partial \sigma^{2}}-(\frac{\partial S_{1}}{\partial
\sigma})^2\right)+a^3 U(\sigma) = 0.\ee Now in order to identify
this equation (36) with the functional Schr\"{o}dinger equation
for the matter field $\sigma$, let us define a function,
\be
f(a,\phi_{1},\phi_{2},\sigma) =
D(a,\phi_{1},\phi_{2})\exp(\frac{i}{\hbar}S_{1}).\ee In view of
(37), equation (36) reduces to
\[ i\hbar\left[\frac{1}{\sqrt{12}a^2}\frac{\partial S_{0}}{\partial \phi_{2}}
\frac{\partial}{\partial a}-\frac{2V_{2}'}{a^3 V_{1}''}\frac{\partial S_{0}}{\partial \phi_{2}}
\frac{\partial}{\partial \phi_{1}}+\left(\frac{1}{\sqrt{12}a^2}
\frac{\partial S_{0}}{\partial a}-\frac{2V_{2}'}{a^3 V_{1}''}\frac{\partial S_{0}}
{\partial \phi_{1}}-\frac{2\phi_{1}+\sqrt 3\phi_{2}-1}{a^3}\frac{\partial S_{0}}
{\partial \phi_{2}}\right)\frac{\partial}{\partial \phi_{2}}\right]f\]\be
= \left[-\frac{\hbar^2}{2a^3}\frac{\partial^2}{\partial\sigma^2}+a^3 U(\sigma)\right]f,\ee
Provided, the function $D(a,\phi_{i})$ satisfies the following equation,
\[ [-\frac{1}{\sqrt{12}a^2}\frac{\partial^2 S_{0}}{\partial a\partial \phi_{2}}
+\frac{2V_{2}'}{a^3 V_{1}''}\frac{\partial^2
S_{0}}{\partial\phi_{1}\partial \phi_{2}} +\frac{2\phi_{1}+\sqrt
3\phi_{2}-1}{2a^3}\frac{\partial^2 S_{0}}{\partial \phi_{2}^2}
+\frac{1}{\sqrt{12} a^2}(\frac{\partial S_{0}}{\partial
a}\frac{\partial}{\partial \phi_{2}} +\frac{\partial
S_{0}}{\partial \phi_{2}}\frac{\partial}{\partial a})\]\be
-\frac{2V_{2}'}{a^3 V_{1}''}(\frac{\partial S_{0}}{\partial
\phi_{1}} \frac{\partial}{\partial \phi_{2}}+\frac{\partial
S_{0}}{\partial \phi_{2}} \frac{\partial }{\partial
\phi_{1}})-\frac{2\phi_{1}+\sqrt 3\phi_{2}-1}{a^3} (\frac{\partial
S_{0}}{\partial \phi_{2}})\frac{\partial}{\partial
\phi_{2}}]D(a,\phi_{i}) = 0.\ee Since the right hand side of
equation (38) is the quantum Hamiltonian operator for the scalar
field $\sigma$ operating on function $f$, in the background of
curved space time, so it is clear that equation (38) is the
functional Schr\"{o}dinger equation for the field $\sigma$,
propagating in the background of curved space time, under the same
choice of the time parameter (23). Further, under the same choice
of time parameter (23), equation (39) reduces to, \be
\frac{1}{D}\frac{\partial D}{\partial t} =
\frac{1}{\sqrt{12}a^2}\frac{\partial^2 S_{0}}{\partial
a\partial\phi_{2}} -\frac{2V_{2}'}{a^3 V_{1}''}\frac{\partial^2
S_{0}}{\partial\phi_{1}\partial\phi_{2}}-\left(\frac{2\phi_{1}+\sqrt
3\phi_{2}-1}{2a^3}\right)\frac{\partial^2 S_{0}}{\partial\phi^2}. \ee
Hence, upto this order of approximation, we have, \be \Psi =
\frac{1}{D}\exp(\frac{i}{\hbar}MS_{0})f, \ee where, $S_{0}$ is
obtainable from equation (35), $f$ from (38) and $D$ from (39) or
(40). So at the end we observe that under the same choice of time
parameter (23), one can parameterize the classical trajectories by
reproducing classical source free gravitational equation (6) from
the source free Hamilton-Jacobi equation (35) at one hand, while
the $M^0$ order term (36) can be identified with the functional
Schr\"{o}dinger equation (38) for the field $\sigma$ propagating
in the background of curved space time, on the other. This time
parametrization is unique since equation (38) reduces to the
functional Schr\"{o}dinger equation only under the choice of the
real time parameter (23) leaving no arbitrariness. This choice is
correct since it fulfills both the requirements of obtaining
functional Schr\"{o}dinger equation and parametrization of the
classical trajectories.
\section{Concluding remarks:}
In a series of works, Brandenberger, Mukhanov, Sornborger and others \cite{mb}, \cite{bms}, \cite{rb} and \cite{rre} constructed a nonsingular gravitational action by invoking limiting curvature hypothesis (LCH) and studied its classical aspects. The action contains a couple of scalar fields, one of which ($\phi_{2}$) is required to guarantee nonsingular solutions, while the other ($\phi_{1}$) binds the curvature invariants. The corresponding potentials are chosen in such a manner that for $|\phi_{i}| \ll 1$, Einstein's theory is recovered at one hand and their asymptotic behaviour should be such that LCH may be realized on the other. In the present work, we have studied the quantum cosmological aspect of the theory and made a connection of the quantum cosmological equation with the Hamilton-Jacobi equation of the classical theory. The answer to the question that, why it is at all necessary to quantize a nonsingular Universe model, lies in fact that some new results have been unveiled in the process and a non-trivial technique of parameterizing the classical trajectories has been found. Summarily, the results are,\\
1.)~ Introduction of the scalar field $\phi_{1}$ in the action (1), only introduces yet another constraint equation (other than the Hamiltonian constraint). Thus it is essentially a classically constrained theory of gravity and so one has to deal with such constraints critically, which we have done using Dirac's algorithm.\\
2.)~  In the absence of the scalar fields, the classical field equations
reduce to those corresponding to standard theory of gravity. However,
the Hamiltonian obtained after analyzing the constraint, does not give
Einstein's theory back. Likewise, the Wheeler-deWitt equation is free
from first derivative terms $\frac{\partial \psi}{\partial a}$, and thus
is quite different from those obtained in view of standard and other
 nonstandard (nonminimally coupled) theories of gravity.\\
3.) The wave function is real for $k = \pm 1$, unlike the situation
encountered in standard gravity theory (see appendix), where, $k = +1$
leads to a complex wave function.\\
4.) ~The semiclassical time parameter required to parameterize classical
trajectories can be found in a unique way only by using the technique of
Lagrange undetermined multiplier in the Lagrangian constrained dynamics, rather than
using Hamiltonian formalism.
\section{Appendix}
\textbf{Paramerizing Einstein's gravity with a minimally coupled scalar field with semiclassical time:}\\
In the appendix we shall recapitulate the standard technique \cite{ck}, \cite{tp}, \cite{tps},
to parameterize classical trajectories of the Hamilton-Jacobi
equation, with some appropriate choice of semiclassical time
parameter, corresponding to Einstein's gravity with a minimally
coupled scalar field, and to find the functional Schr\"{o}dinger equation.\\
The action for Einstein's gravity with a source term - a minimally
coupled scalar field $\phi$, (with $M = \frac{3\pi}{2G}$, which is different from $M = \frac{\pi}{8G}$, used in the text) is,
\[A = M\int[-\frac{1}{2}a\dot a^2 +k\frac{a}{2}-\frac{1}{M}(\frac{1}{2}\dot\phi^2 -V(\phi))a^3]dt,\]
apart from a surface term. The ($^0_{0}$) component of Einstein's
equation is \be -\frac{M}{2}\left(\frac{\dot
a^2}{a^2}+\frac{k}{a^2}\right)+\frac{1}{2}\dot\phi^2 + V(\phi) =
0,\ee which corresponds to the Hamilton constraint equation, \be
-\frac{1}{2M}\frac{p_{a}^2}{a}+\frac{p_{\phi}^2}{2 a^3} -
\frac{M}{2} k a + a^3 V(\phi) = 0.\ee Thus, the corresponding
Wheeler-deWitt equation is \be
\left[\frac{\hbar^2}{2M}(\frac{\partial^2}{\partial
a^2}+\frac{p}{a}\frac{\partial}{\partial
a}-\frac{M}{a^2}\frac{\partial^2}{\partial\phi^2}]-\frac{M}{2}k
a^2 +a^4 V(\phi)\right]|\Psi> = 0,\ee where, $p$ takes care of
some of the operator ordering ambiguities.
\subsection{Expansion with power series of Planck's constant $\hbar$}
In the standard WKB approximation, the wave functional $\Psi$ is
expressed as $\Psi(a,\phi) = \exp{[\frac{i}{\hbar}S(a, \phi)]}$,
and the functional $S(a, \phi)$ is expanded in the power series of
the Planck's constant $\hbar$ as, $S = S_{0} + \hbar S_{1}
+\hbar^2 S_{2} + \cdots$ etc. Substituting all these in the
Wheeler-deWitt equation (44), and collecting terms independent of
$\hbar$, one obtains, \be-\frac{1}{2M}\left(\frac{\partial
S_{0}}{\partial a}\right)^2 + \frac{1}{2a^2}\left(\frac{\partial
S_{0}}{\partial \phi}\right)^2 -\frac{M}{2}k a^2 + a^4 V(\phi) =
0.\ee This is Einstein-Hamilton-Jacobi (EHJ) equation, which is
essentially equation (43), if one identifies $p_{a}$ with
$\frac{\partial S_{0}}{\partial a}$ and $p_{\phi}$ with
$\frac{\partial S_{0}}{\partial \phi}$. In order to parameterize
classical trajectories, WKB time parameter is chosen in the
following manner, \be\frac{\partial S_{0}}{\partial a} = p_{a} =
-M\dot a a,\;\;or,\;\;\frac{\partial a}{\partial t} =
-\frac{1}{Ma}\frac{\partial S_{0}}{\partial a},\;\;
ie.,\;\;\frac{\partial}{\partial t}|_{a} =
-\frac{1}{Ma}\frac{\partial S_{0}}{\partial
a}\frac{\partial}{\partial a},\ee and \be\frac{\partial
S_{0}}{\partial \phi} = p_{\phi} = a^3 \dot\phi,
\;\;or,\;\;\frac{\partial\phi}{\partial t} =
\frac{1}{a^3}\frac{\partial S_{0}}{\partial
\phi},\;\;ie.,\;\;\frac{\partial}{\partial t}|_{\phi} =
\frac{1}{a^3}\frac{\partial S_{0}}{\partial
\phi}\frac{\partial}{\partial\phi}.\ee The classical trajectories
are parameterized by a time, which is a linear combination of (46)
and (47). Thus, \be \frac{\partial}{\partial t} =
-\frac{1}{Ma}\frac{\partial S_{0}}{\partial
a}\frac{\partial}{\partial a}+ \frac{1}{a^3}\frac{\partial
S_{0}}{\partial \phi}\frac{\partial}{\partial\phi}.\ee Under this
choice of time parameter (48), equation (45) produces Einstein's
equation (42). This has been shown by Kiefer \cite{ck}, taking, $a
= \exp\alpha$ and using the unit $M = 1$. It is important to note
that the choice of the semiclassical time parameter does not
involve operator ordering ambiguity. Equation (45) can in
principle be solved for $S_{0}$ and to this order of approximation
one obtains $\Psi = \exp{(\frac{i}{\hbar} S_{0})}$. Higher order
terms of $\hbar$ produce fluctuations around the classical
trajectory $S_{0}$ and hence introduce corrections to the
wave-functional $\Psi$.
\subsection{Expansion with power series of Planck's mass $M$}
In the present work, we have expanded $S(a, \phi )$ in the power
series of Planck's mass, instead. The advantage is that, one can
obtain a functional Schr\"{o}dinger equation directly from it.
Substituting, $S = M S_{0} + S_{1} + M^{-1} S_{2} + \cdots etc.,$
in the Wheeler-deWitt equation (44) and collecting terms in
different orders in $M$, we obtain for $M^2$ order,
$\frac{\partial S_{0}}{\partial\phi} = 0$. It implies that $S_{0}$
is purely a functional of gravitational field. $M^1$ order term
gives the source free Einstein-Hamilton-Jacobi (EHJ) equation, \be
\frac{a^2}{2}\left(\frac{\partial S_{0}}{\partial a}\right)^2 +
\frac{1}{2} k a^4 = 0.\ee Now, in the absence of the matter field,
equations (42) and (43) reduce to \be \dot a^2+k =
0,\;\; and, \;\; \frac{a^2}{2M}p_{a}^2 + \frac{M}{2}k a^4 = 0, \ee
respectively. Hence the EHJ equation (49) is essentially the
vacuum Einstein equation (50), under the identification $p_{a} =
M\frac{\partial S_{0}}{\partial a}$. To parameterize classical
trajectories, we identify, \be M\frac{\partial S_{0}}{\partial a}
= p_{a} =- Ma\dot a,\;\;ie.,\;\; \frac{\partial}{\partial t} =
-\frac{1}{a}\frac{\partial S_{0}}{\partial
a}\frac{\partial}{\partial a}.\ee Further in view of equation
(49), we have, \be \frac{\partial S_{0}}{\partial a} = \pm i\sqrt
k a,\;\;ie.,\;\;\frac{\partial}{\partial t} = \pm i\sqrt k
\frac{\partial}{\partial a}.\ee Under the choice of the time
parameter (52), equation (49) reduces to the vacuum Einstein's
equation (50). Hence, expansion of $S$ in the power series of $M$
decouples gravity from the source and leaves it to behave
classically. Equation (49) can be solved for $S_{0}$, and up to
this order of approximation,

\be\Psi(a) = \exp{[\pm \frac{M}{2\hbar} \sqrt k a^2]}.\ee
Thus the wave-functional is well behaved at the classical singularity
$a \rightarrow 0$.
The next order of approximation yields the following equation, \be
\frac{i\hbar}{2}\left(\frac{\partial^2 S_{0}}{\partial a^2}
+\frac{p}{a}\frac{\partial S_{0}}{\partial a}-\frac{1}{a^2}
\frac{\partial^2 S_{1}}{\partial \phi^2}\right)-\frac{\partial
S_{0}} {\partial a}\frac{\partial S_{1}}{\partial a}+\frac{1}{2
a^2} \left(\frac{\partial S_{1}}{\partial \phi}\right)^2 + a^4
V(\phi) = 0\ee where, $p$ takes care of some of the factor
ordering ambiguities. Now choosing a function, $f(a, \phi) = D(a)
\exp{[\frac{iS_{1}}{\hbar}]}$ and substituting $\frac{\partial
S_{1}}{\partial a}, \frac{\partial S_{1}}{\partial \phi}$ and
$\frac{\partial^2 S_{1}}{\partial \phi^2}$ from it in the above
equation (54), one obtains, \be
-\frac{i\hbar}{a}\left(\frac{\partial S_{0}}{\partial
a}\right)\frac{\partial}{\partial a}f(a, \phi) =
\left(-\frac{\hbar^2}{2a^3}\frac{\partial^2}{\partial\phi^2} + a^3
V(\phi)\right) f(a, \phi),\ee provided, $D(a)$ satisfies the following equation,
\be \frac{\partial S_{0}}{\partial a}\frac{\partial
D(a)}{\partial a} - \frac{1}{2}\left(\frac{\partial^2
S_{0}}{\partial a^2} + \frac{p}{a}\frac{\partial S_{0}}{\partial
a}\right) D(a) = 0.\ee Equation (55) can be identified with
Tomonaga-Schwinger equation, which is essentially the functional
Schr\"{o}dinger equation for the mater field propagating in the
background of curved space-time, if the semi-classical time
operator is identified with (51). The right hand side of equation
(55) represents quantum Hamilton operator $\hat{H}_{m}$ for the
matter field operating on the function $f(a, \phi)$. Thus equation
(55) reduces to \be i\hbar \frac{\partial f}{\partial t} =
\hat{E}f(a, \phi) = \hat{H}_{m}f(a, \phi).\ee Equation (56) can be
solved for $D(a)$ as, \be D(a) = m
\sqrt{S_{0}'}~a^{\frac{p}{2}}.\ee  Substituting $S_{0}' = \pm i\sqrt k a$,
in equation (56), one obtains,
\be D(a) = m a^{(\frac{1+p}{2})},\ee $m$ being the
constant of integration.
Finally, to this order of approximation, the wave-functional takes
the form, \be \Psi = \exp{[\frac{i}{\hbar}(MS_{0} + S_{1})]} =
\frac{a^{-(\frac{p+1}{2})}}{m}\exp{[\pm\frac{M}{2\hbar}\sqrt k
a^2]f(a, \phi)}.\ee Equation (55) can in principle be solved for
$f(a, \phi)$ and thus explicit form of the wave-functional can be
obtained.\\
Now to make semiclassical approximation to the
functional Schr\"{o}dinger equation (55), let us
express $f(a, \phi) = \exp{[\frac{i}{\hbar} A(a, \phi)]}$,
expand $A$ in the power series of $\hbar$ as, $A(a, \phi) =
A_{0} + \hbar A_{1} + \hbar^2 A_{2} + \cdots$, substitute it
in equation (55) and collect terms independent of $\hbar$, to
get, \be -\frac{1}{a}\frac{\partial S_{0}}{\partial
a}\frac{\partial A_{0}} {\partial a} =
\frac{1}{2a^3}\left(\frac{\partial A_{0}}{\partial\phi}\right)^2 +
a^3 V(\phi).\ee Since we have already identified
$-\frac{1}{a}\frac{\partial S_{0}} {\partial
a}\frac{\partial}{\partial a}$ with the semiclassical time
parameter $\frac{\partial}{\partial t}$, in equation (52), so the
above equation takes the form, \be \frac{\partial A_{0}}{\partial
t} - \frac{1}{2a^3} \left(\frac{\partial A_{0}}{\partial
\phi}\right)^2 - a^3 V(\phi) = 0.\ee Further, if one identifies
$\frac{\partial A_{0}}{\partial \phi}$ with $p_{\phi}$, it transpires that equation (62) is the Hamiltonian for the matter
field in the background of curved space time, corresponding to
the classical action,
$\int d^4 x\sqrt{-g} L_{m} = 2\pi^2 \int [\frac{1}{2}\dot\phi^2 - V(\phi)]a^3 ~dt,$ in the isotropic and homogeneous space-time under consideration.
Now, since $p_{\phi} = a^3 \dot\phi = \frac{\partial A_{0}}{\partial\phi}$, so
$$\frac{\partial}{\partial t} = \frac{1}{a^3}\frac{\partial A_{0}}{\partial\phi}\frac{\partial}{\partial\phi},$$
and the suppressed part of the WKB time parameter appearing in
equation (48) reappears. This when combined with equation (51), one finally
obtains the semiclassical time parameter as,

\be \frac{\partial}{\partial t} =
-\frac{1}{a}\frac{\partial S_{0}} {\partial
a}\frac{\partial}{\partial a} + \frac{1}{a^3} \frac{\partial
A_{0}}{\partial \phi}\frac{\partial}{\partial\phi}.\ee
\noindent
{\bf{Acknowledgement :}} I would like to thank prof. R. Brandenberger and the reviewers
for going through the manuscript and rendering some valuable comments and suggestions.\\

\end{document}